\let\Xdocument\document
\documentclass{iau}
\let\document\Xdocument

\usepackage{amsmath}
\usepackage{graphicx}
\usepackage{multirow}

\begin{document}

\lefttitle{Bhardwaj A.}
\righttitle{Near-infrared Period--Luminosity--Metallicity relations for Cepheid and RR Lyrae variables}

\jnlPage{1}{7}
\jnlDoiYr{2023}
\doival{10.1017/xxxxx}

\aopheadtitle{Proceedings IAU Symposium}
\editors{R. de Grijs,  P. Whitelock \&  M. Catelan, eds.}

\title{Period--Luminosity--Metallicity relations for Classical Pulsators at Near-infrared Wavelengths}

\author{Anupam Bhardwaj}
\affiliation{INAF - Osservatorio Astronomico di Capodimonte, Via Moiariello 16, 80131, Napoli, Italy}

\begin{abstract}
Classical pulsating stars such as Cepheid and RR Lyrae variables exhibit well-defined Period--Luminosity relations at near-infrared wavelengths. Despite their extensive use as stellar standard candles, the effects of metallicity on Period--Luminosity relations for these pulsating variables, and in turn, on possible biases in distance determinations, are not well understood. We present ongoing efforts in determining accurate and precise metallicity coefficients of Period--Luminosity-Metallicity relations for classical pulsators at near-infrared wavelengths. For Cepheids, it is crucial to obtain a homogeneous sample of photometric light curves and high-resolution spectra for a wide range of metallicities to empirically determine metallicity coefficient and reconcile differences with the predictions of the theoretical models. For RR Lyrae variables, using their host globular clusters covering a wide range of metallicities, we determined the most precise metallicity coefficient at near-infrared wavelengths, which is in excellent agreement with the predictions of the horizontal branch evolution and stellar pulsation models.

\end{abstract}


\maketitle

\section{Introduction}

Cepheid and RR Lyrae variables are the most popular subclasses of radially pulsating stars and are often referred as classical pulsators. These variables are robust stellar standard candles for astronomical distance measurements because their luminosities are strongly correlated with their pulsation periods \citep[see the reviews by][]{subramanian2017, bhardwaj2020}. The Period-Luminosity (PL) relation or the ``Leavitt Law'' is at the basis of the extragalactic distance ladder used to determine the Hubble constant or the present expansion rate of the Universe \citep{freedman2001, riess2022}. At present, there is a discord between the Hubble constant values based on the traditional Classical Cepheid-Supernova distance ladder \citep{riess2022} and those from the {\it Planck} mission \citep{planck2020}. The discrepancy in these two measurements of the expansion rate at two different ends of the Universe hint at missing or new physics in the standard cosmological model. Since the PL relation of Classical Cepheid variables plays a vital role in the local determination of the Hubble constant, it is now crucial to explore all possible sources of systematic uncertainties in the calibration of these relations and improve the accuracy and the precision of the extragalactic distance ladder. 

Despite the extensive use of PL relations for distance measurements, their universality in different stellar environments has not been established and is often debated in the literature \citep{bhardwaj2020}. When approaching a percent-level precision, the impact of chemical compositions, topology of the instability strip, and the mass-luminosity relations obeyed by classical pulsators on PL relations can also become significant \citep{bono2010, marconi2015}. Some of these evolutionary effects become negligible at near-infrared (NIR) wavelengths because of the narrow width of the instability strip and small temperature variations. NIR observations also offer advantages over optical bands for distance measurements because the PL relations are tighter at these longer wavelengths and extinction effects are significantly smaller. Furthermore, smaller variability amplitudes in NIR lead to more accurate mean magnitude determinations even with a small number of observations or sparsely sampled light curves. The absolute calibration of NIR PL relations for classical pulsators are now crucial in view of the upcoming large ground-based observational facilities and space telescopes, which will predominantly operate at these longer wavelengths. 

Even at NIR wavelengths, one of the main sources of residual systematic uncertainty on the Classical Cepheid based distance scale is due to metallicity effects on PL relations \citep{ripepi2021, riess2022}. While there is no consensus on the exact dependence of metallicity, it is expected to affect both the slope and zero-point of the PL relations \citep{bono2010}. \citet{breuval2022} provided a compilation of theoretical and empirical estimates of the metallicity coefficient (commonly referred as $\gamma$ in the literature) of Classical Cepheid PL relations over the past two decades. It is interesting to note that the metallicity term of the $K_s$-band Period-Luminosity-Metallicity (PLZ) relation varies significantly between $-0.45$ and $+0.02$ mag/dex in different empirical studies, and thus, even the sign of the metallicity dependence is debated. However, more recently the values of $\gamma$ are converging between $-0.4$ and $-0.2$ mag/dex in NIR bands. In the case of Type II Cepheids, both empirical and theoretical calibrations predict a small or negligible dependence in limited studies at NIR wavelengths \citep{matsunaga2006, das2021}. Recently, \citet{wielgorski2022} found a relatively larger dependence of the order of $-0.2$ mag/dex in $JHK_s$ band PL relations albeit using a small sample of Type II Cepheid variables. Similar to Cepheids, the distance measurements based on RR Lyrae stars will also result in large systematic biases if the metallicity term on their PL relation is not taken into account \citep{catelan2004, marconi2015, bhardwaj2023}. Table~\ref{tab:rrl} lists the empirical and theoretical estimates of metallicity coefficient in NIR filters in different stellar environments. It is evident that the sign of metallicity dependence is well established for RR Lyrae stars such that metal-rich stars are fainter. Moreover, the metallicity coefficient in NIR bands seems to converge around 0.2 mag/dex in both empirical and theoretical studies. 

 \begin{table}[h!]
 \begin{center}
 \caption{Metallicity coefficients of RR Lyrae PLZ relations in NIR bands.}\label{tab:rrl}
 {\small\begin{tabular}{@{\extracolsep{\fill}}lccllll}
    \midrule
     & $N_{\textrm{RRL}}$& $\Delta$[Fe/H]& $J$ & $H$ & $K_s$ & Ref.\\
     \midrule
GC + MW & 1310 &    $2.0$ & \bf $0.20\pm0.02$     & \bf $0.19\pm0.01$       & \bf $0.18\pm0.01$ & \citet{bhardwaj2023}\\
MW   &   448    &   $2.8$ & ---   & ---    & \bf $0.10\pm0.03$   &  \citet{muhie2021}\\
GC + MW    &   403    &   $2.7$ & ---   & ---   & \bf $0.17\pm0.01$   &  \citet{bhardwaj2021}\\
MW &  55    &   $2.5$ & \bf $0.20\pm0.03$     & \bf $0.17\pm0.03$       & \bf $0.17\pm0.03$   &  \citet{neeley2019}\\
$\omega$ Cen &  55    &   $1.0$ & \bf $0.15\pm0.02$     & \bf $0.13\pm0.02$       & \bf $0.15\pm0.02$   &  \citet{braga2018}\\
$\omega$ Cen & 64   &   $1.0$ & \bf $0.15\pm0.03$     & -      & \bf $0.14\pm0.02$   &  \citet{navarrete2017}\\
Theory   &   ---    &   $--$ &\bf $0.18\pm0.01$     & \bf $0.18\pm0.01$    & \bf $0.18\pm0.01$   &  \citet{marconi2015}\\
MW   &   23    &   $2.6$ & ---   & ---    & \bf $0.07\pm0.04$   &  \citet{muraveva2015}\\
LMC  &  70     &   $1.5$ & ---   & ---    & \bf $0.03\pm0.07$   &  \citet{muraveva2015}\\
MW   &   403    &   $2.0$ & ---   & ---    & \bf $0.09\pm0.03$   &  \citet{dambis2013}\\
LMC   &   50    &   $1.3$ & ---   & ---    & \bf $0.05\pm0.07$   &  \citet{borissova2009}\\
GC   &   538    &   $2.0$ & ---   & ---    & \bf $0.08\pm0.11$   &  \citet{sollima2008}\\
Theory  &   ---    &   $--$ & \bf $0.19\pm-$     & \bf $0.18\pm-$       & \bf $0.18\pm-$   &  \citet{catelan2004}\\
MW   &   27    &   $2.0$ & ---   & ---    & \bf $0.23\pm0.01$   &  \citet{bono2003}\\
Theory   &   ---    &   $--$ & ---   & ---    & \bf $0.17\pm-$   &  \citet{bono2001}\\
    \midrule
    \end{tabular}}
 \end{center}
\small{\textit{Notes}: [1] RR Lyrae host stellar systems in column 1 are: GC - globular clusters, MW- Milky Way, LMC - Large Magellanic Cloud.}
\end{table}

In the following subsections, we will discuss our recent and ongoing efforts to improve the quantification of metallicity dependence on PL relations for Cepheid and RR Lyrae stars. 

\section{Classical or Type I Cepheid variables}

Until a few years back, a precise determination of metallicity coefficient of Classical Cepheid PLZ relations was hampered by the lack of accurate distances to Galactic Cepheids. High precision parallaxes for thousands of Classical Cepheids are now being provided by the {\it Gaia} astrometric mission along with their photometric light curves \citep{ripepi2022}. However, {\it Gaia} does not provide high-resolution spectroscopic abundances and NIR light curves of classical pulsators, which are equally important for deriving precise PLZ relations. 
  
    \begin{figure*}[!h]
    \includegraphics[scale=.76]{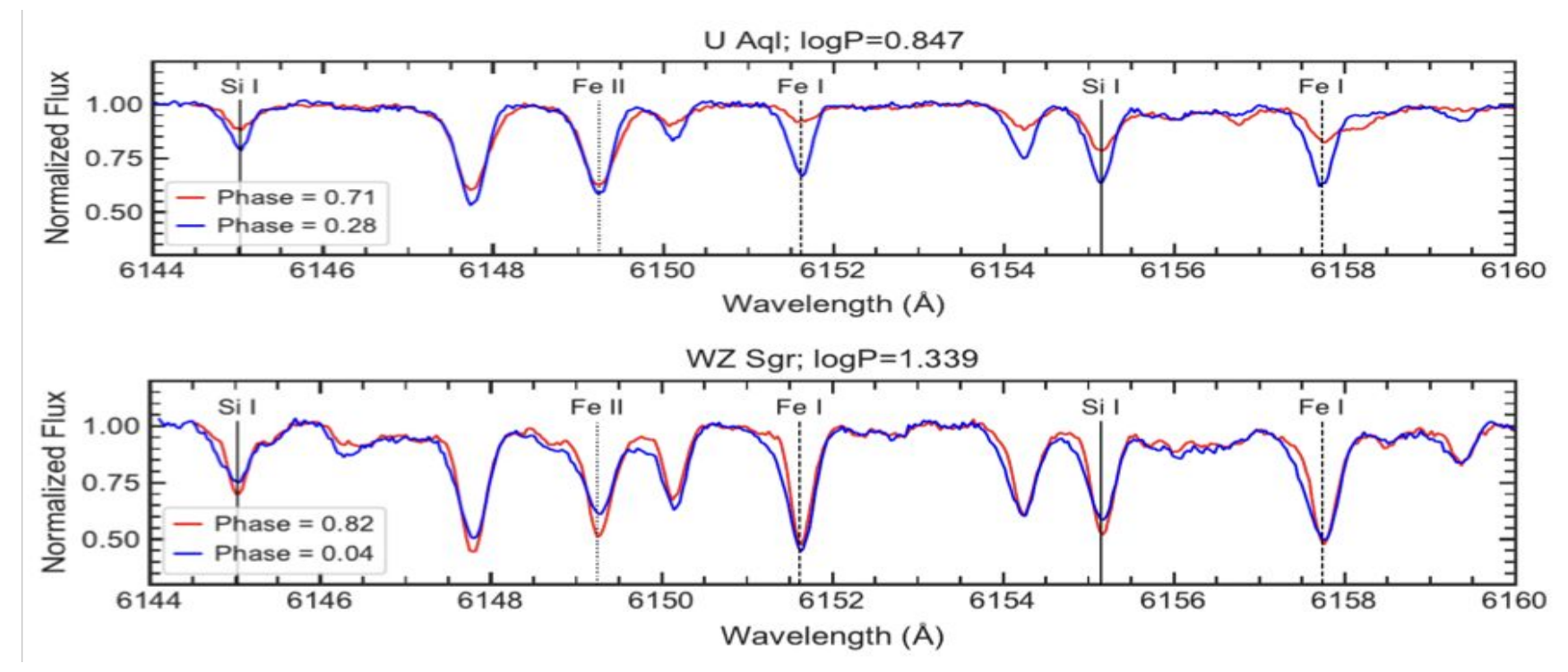}
    \caption{Optical spectra of a short-period (top) and a long-period (bottom) Milky Way Cepheid standard at two different phases along their pulsation cycles. Iron lines are marked as dotted lines that are used to determine spectroscopic metallicities ([Fe/H]).}
    \label{fig:cep_spec}
    \end{figure*} 
  
\subsection{High-resolution spectroscopic metallicities}

High-resolution spectroscopic metallicities of Classical Cepheid variables are available in the literature for small samples of stars in a single study, which were then used to homogenize larger samples of literature [Fe/H] measurements in different metallicity scales \citep[e.g.,][]{genovali2014, luck2018}. However, Classical Cepheids with metallicities measured using high-resolution spectra were mainly restricted to the solar neighbourhood and there are a very few variables with low metallicities ([Fe/H]$<-0.4$~dex). Since the solar neighbourhood Classical Cepheids exhibit a rather narrow range of [Fe/H] ($\sigma=0.3$~dex), it is difficult to properly quantify the metallicity term of the PLZ relations. In the latest determination of the Hubble constant, 75 Milky Way Cepheid standards were used to calibrate the PLZ relation for the extragalactic distance ladder along with independent anchors such as LMC, and NGC 4258 \citep{riess2022}. These 75 Milky Way Cepheid standards were carefully selected to be in less crowded and less extincted regions in the Galaxy and exhibit a [Fe/H] range of 0.7 dex. The spectroscopic [Fe/H] values for these Cepheids were adopted from \citet{groenewegen2018} based on the compilation of \citet{genovali2015}. Homogeneous high-resolution spectroscopy of these 75 Milky Way Cepheid standards is crucial to properly quantify the metallicity term and measure its impact on the distance scale and the Hubble constant.  

    \begin{figure*}
    \includegraphics[scale=.55]{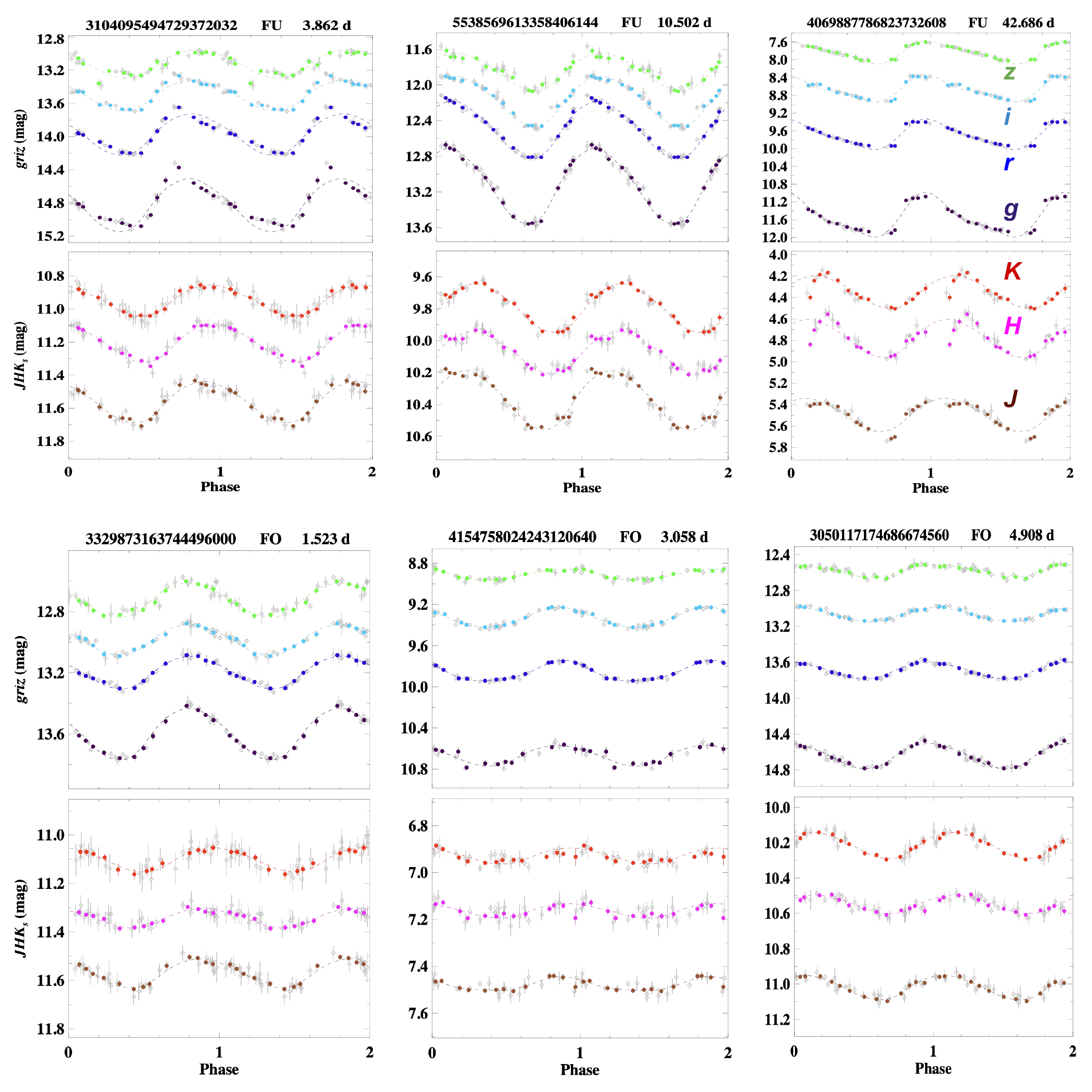}
    \caption{Example light curves of fundamental (FU) and first-overtone (FO) mode Cepheids with different periods in our sample. The grey points in each panel show all the data points while filled circles represent phase averaged light curves. The best-fitted sinusoidal templates are also overplotted. The {\it Gaia} DR3 Source ID, pulsation mode, and periods are listed on the top of each panel. }
    \label{fig:cep_lc}
  \end{figure*}
  
We are obtaining high-resolution spectra of Milky Way Cepheid standards using the ESPaDOnS instrument ($R\sim81,000$) at the Canada-France-Hawaii Telescope (CFHT). In addition to optical spectra, we are also collecting high-resolution  ($R\sim45,000$) $HK_s$ band spectra for these Classical Cepheids for the first-time using IGRINS instrument at Gemini-South Telescope. Multi-epoch spectra for 42 out of 75 Classical Cepheids have already been obtained and analysed to derive [Fe/H] values. Fig.~\ref{fig:cep_spec} displays the spectra of two Cepheid standards at two different phases along their pulsation cycle. The iron abundances were measured using the same methodology as discussed in \citet{ripepi2021} and the interested readers are referred to this paper for more details. 
The preliminary analysis with derived [Fe/H] values suggests an excellent agreement with the [Fe/H] values from \citet{luck2018}, which are also based on multi-epoch spectra of Classical Cepheid variables. Our spectroscopic [Fe/H] values for 42 Milky Way Cepheids were combined with the remaining 35 Cepheids having metallicities from \citet{luck2018} to obtain a homogenized sample of 75 stars with new [Fe/H] measurements. Multiband photometric light curves for majority of these 75 Cepheids were analysed by \citet{bhardwaj2015, bhardwaj2017} using a Fourier decomposition technique. Therefore, accurate mean magnitudes in $VIJHK_s$ bands are available, and together with high-resolution [Fe/H] measurements and precise {\it Gaia} parallaxes, our sample will be used to derive PLZ relations for these Milky Way standard Classical Cepheid variables. 

 At the same time, high-resolution spectroscopic metallicities for more metal-poor Classical Cepheids are being obtained as part of the Cepheid Metallicity in the Leavitt Law (C-MetaLL) survey \citep{ripepi2021}. This program aims to enlarge the metallicity range of Classical Cepheids by observing more distant targets in the Galactic disk and anti-center directions using HARPS instrument at Telescopio Nazionale Galileo (TNG) and UVES instrument at Very Large Telescope (VLT). \citet{ripepi2021} found a rather large metallicity coefficient in NIR bands of the order of $\sim-0.4$ mag/dex. This is in contradiction to \citet{riess2021} value of $\sim-0.2$~mag in the $H$-band Wesenheit relation, which was determined using 75 Milky Way Cepheids in the solar vicinity. Note that the Wesenheit magnitudes \citep[see][for details]{madore1982, inno2015, bhardwaj2016a} are by construction reddening-independent. 
 Most studies which either used a small sample of Milky Way Cepheids in the solar neighbourhood or combined those with Classical Cepheids in the Magellanic Clouds to enlarge the metallicity range, have found a metallicity coefficient of the order of $-0.2$ mag/dex in NIR bands \citep[e.g.][]{gieren2018, breuval2022}. Therefore, increasing the metallicity range using more than 110 published C-MetaLL target Classical Cepheids is crucial to better constrain the metallicity coefficient of the PLZ relation. However, since these targets are more distant, there are no NIR light curves for these stars in the literature and their latest {\it Gaia} parallaxes also exhibit larger uncertainties.

 \subsection{Homogeneous optical and NIR photometric light curves}

 Multiband light curves of Classical Cepheid variables are useful to not only obtain precise and accurate mean magnitudes, but also determine their reddening values. Optical light curves for Classical Cepheids in traditional Johnson-Cousins filters are available in the literature from large variability surveys (e.g., $VI$ in OGLE or $V$ in ASAS-SN) or from photoelectic observations \citep[e.g.][in $UBVRI$]{berdi2008}. However, no homogeneous light curves in Sloan (optical) and NIR filters are available for most of our target Cepheid variables with spectroscopic metallicities. The calibration of Cepheid PLZ relations in the Sloan-like and NIR filters will be particularly useful in the era of Vera C. Rubin Observatory's Legacy Survey of Space and Time (LSST) and upcoming thirty meter class telescopes operating at infrared wavelengths, respectively. 

We are obtaining multiband ($grizJHK_s$) light curves of Cepheids using the Rapid Eye Mount (REM) telescope. The observations are collected simultaneously in all filters using two different instruments - ROSS (optical) and REMIR (NIR). The observations were scheduled nightly for a given target, if visible, for up to 60 nights within a semester. We have already obtained imaging data for 80 Cepheids and reduced the images using DAOPHOT/ALLFRAME suite of softwares \citep{stetson1996}. The light curves were calibrated in Sloan Digital Sky Survey (SDSS) filters using {\it Gaia} synthetic photometry and in NIR filters using Two Micron All Sky Survey \citep[2MASS,][]{skrutskie2006} observations. Fig.~\ref{fig:cep_lc} displays preliminary light curves of fundamental (FU) and first-overtone (FO) mode Cepheids in our sample. The number of data points per light curve range between 10 and 60, and there are on average 25 epochs. Template-fitted light curves were used to obtain mean-magnitudes for these Classical Cepheid variables and derive preliminary PL relations. Our analysis suggests a large scatter in the PL relations due to larger parallax uncertainties for these distant targets, and will be presented in detail in Bhardwaj et al. (2023, in prep.). Our new homogeneous spectra and multiband photometry of Galactic Classical Cepheids will be useful to derive accurate and precise PLZ relations. 

\section{Type II Cepheids}

Type II Cepheids have never been used as extensively as  Classical Cepheids for distance measurements primarily because these are fainter and less abundant. These old and low-mass stars are further separated in three distinct subclasses of BL Herculis (BL Her), W Virginis (W Vir), and RV Tauri (RV Tau) stars based on their different evolutionary status \citep[see the review by][]{wallerstein2002, bhardwaj2022}. Observationally, the separation of Type II Cepheids in three different subclasses is based on their pulsation periods \citep{soszynski2018}. While these population II stars also follow a tight PL relation in NIR bands, the metallicity dependence on these relations has not been explored in detail in the literature. Similar to Classical Cepheids, the primary reason for not having a well-calibrated PLZ relation is the lack of a homogeneous sample of Type II Cepheids with both high-resolution spectroscopic metallicities and multiband photometry.
Empirically, the best constraints on the metallicity coefficient of Type II Cepheid PL relations in $JHK_s$ bands were provided by \citet{matsunaga2006} using variables in globular clusters. The authors assumed mean-metallicity of the cluster for individual Type II Cepheids and adopted the distances to the clusters based on horizontal branch stars, and combined those with mean $JHK_s$ magnitudes to find a small or negligible metallicity coefficient. Theoretical pulsation models have also predicted no statistically significant metallicity dependence of Type II Cepheid PL relations at NIR wavelengths \citep{di2007, das2021}. The lack of accurate parallaxes in the pre-{\it Gaia} era precluded such empirical studies for Milky Way field Type II Cepheids. More recently, \citet{wielgorski2022} provided NIR time-series photometry for 21 Galactic field Type II Cepheids and obtained a larger metallicity coefficient in NIR filters using 8 stars having high-resolution spectroscopic metallicities. In contrast, \citet{ngeow2022a} found no metallicity dependence on NIR PL relations for Type II Cepheids in globular clusters.

  \begin{figure*}
  \centering
    \includegraphics[scale=.38]{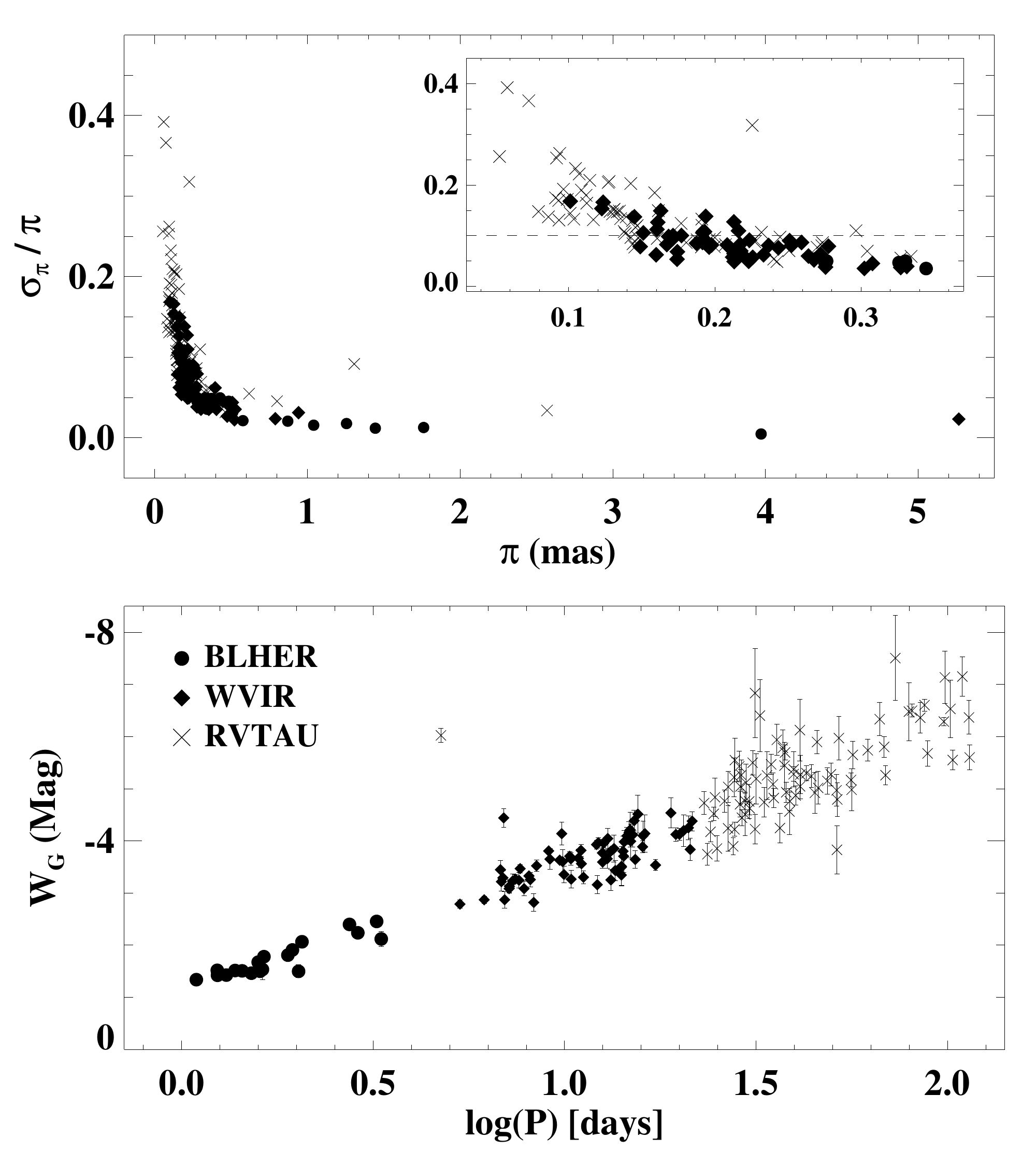}
    \caption{{\it Top:} Parallaxes of 166 Type II Cepheids in {\it Gaia} with spectroscopic metallicities from RVS. It is evident that parallaxes for RV Tau stars exhibit large uncertainties. {\it Bottom:} The Wesenheit magnitudes in {\it Gaia} bands showing tight PW relation for BL Her and W Vir stars and larger scatter for RV Tau variables.}
    \label{fig:t2c_plx}
  \end{figure*}
  
\subsection{Spectroscopic metallicities of Type II Cepheids from {\it Gaia}}

Homogeneous spectroscopic metallicities and NIR time-series photometry for a statistically significant sample of Type II Cepheids is crucial for the empirical calibration of their PLZ relations in NIR bands. This is particularly important now because accurate parallaxes for more than 1600 Type II Cepheids are available in the latest {\it Gaia} data release \citep{ripepi2022}. However, high-resolution spectroscopic metallicities are still limited to small samples of 10-20 Type II Cepheids variables \citep[][]{maas2007, kovtyukh2018}. In addition to precision astrometry, {\it Gaia} also provides medium resolution spectra for bright sources using Radial Velocity Spectrometer (RVS) instrument. We explored {\it Gaia}-RVS spectroscopic metallicities for Type II Cepheids and found 166 variables with both optical light curves and [Fe/H] measurements. This represents the largest sample of Milky Way field Type II Cepheids with homogeneous [Fe/H] determinations.

\begin{figure*}
\centering
\includegraphics[scale=.53]{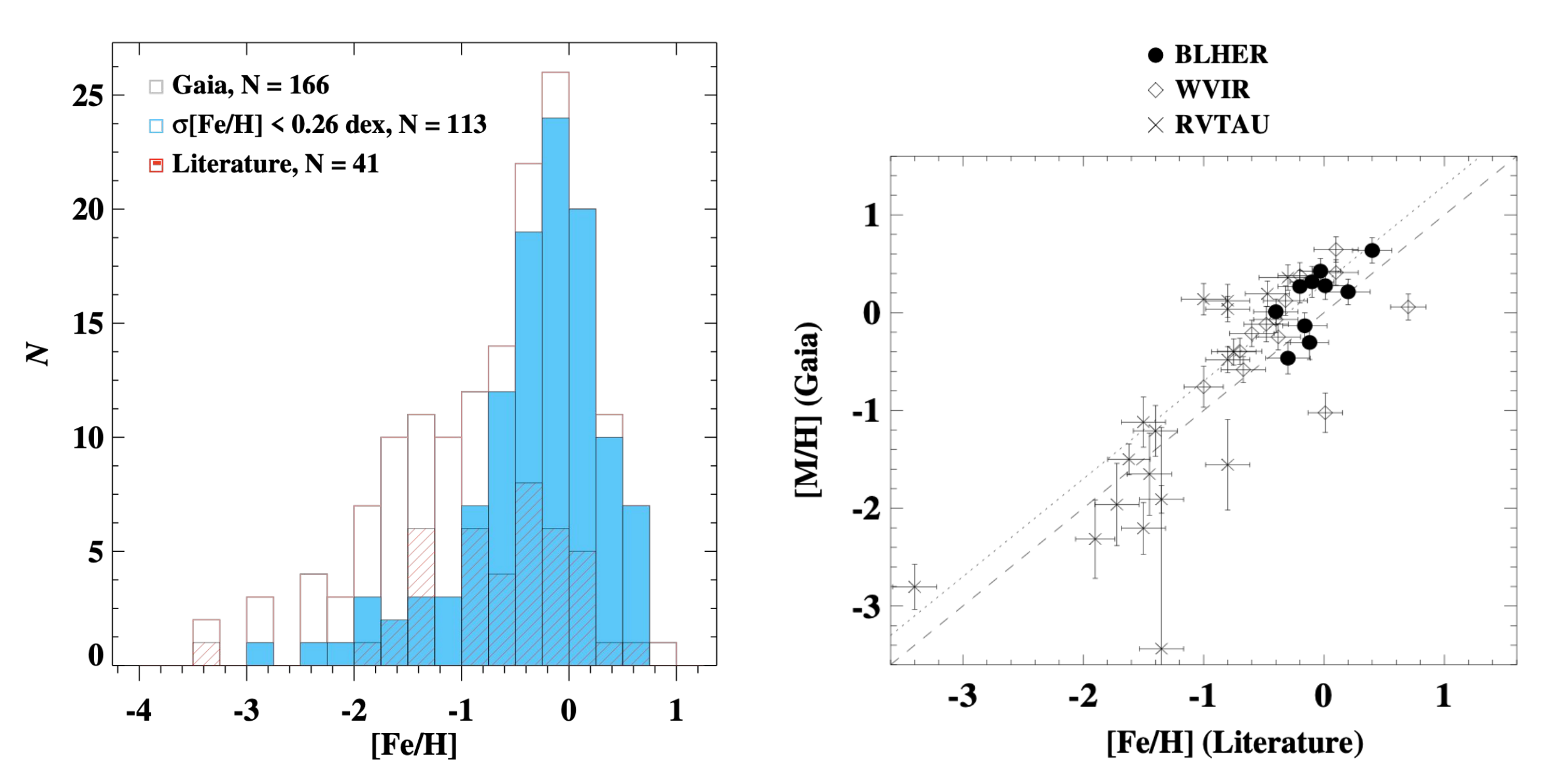}
\caption{~{\it Left:} Histograms of spectroscopic metallicities for {\it Gaia} Type II Cepheids. ~{\it Right:} Comparison of high-resolution spectroscopic metallicities from {\it Gaia}-RVS with those in the literature for common stars.}
\label{fig:t2c_met}
\end{figure*}
  
Fig.~\ref{fig:t2c_plx} displays {\it Gaia} parallaxes and photometry for 166 Type II Cepheids with RVS spectroscopic metallicities. The top panel clearly shows that the parallaxes for RV Tau stars are the most uncertain while BL Her have most accurate and precise parallaxes. This is also evident in the scatter in the Period-Wesenheit (PW) relation in {\it Gaia} bands in the bottom panels. 
Fig.~\ref{fig:t2c_met} displays the histogram of spectroscopic metallicities for all 166 variables. The {\it Gaia}-RVS metallicities have been assigned different quality flags \citep[see][for details]{blanco2022}, and accordingly, we adopted a minimum uncertainty of 0.13 dex corresponding to $1\sigma$ systematics. We considered 113 {\it Gaia}-RVS measurements with acceptable quality flags with a maximum uncertainty of $2\sigma \sim 0.26$ dex. All the {\it Gaia}-RVS measurements were corrected using the recipe suggested by \citet{blanco2022}. For a relative comparison, we compiled 41 common Type II Cepheids with high-resolution spectroscopic metallicities in the literature. The right panel of Fig.~\ref{fig:t2c_met} shows a comparison of {\it Gaia}-RVS spectroscopic metallicities with those available in the literature. We found a systematic offset of $\sim 0.3$ dex between these two sets of [Fe/H] values with measurements for RV Tau stars exhibiting the largest scatter.

  \begin{figure*}
    \includegraphics[scale=.33]{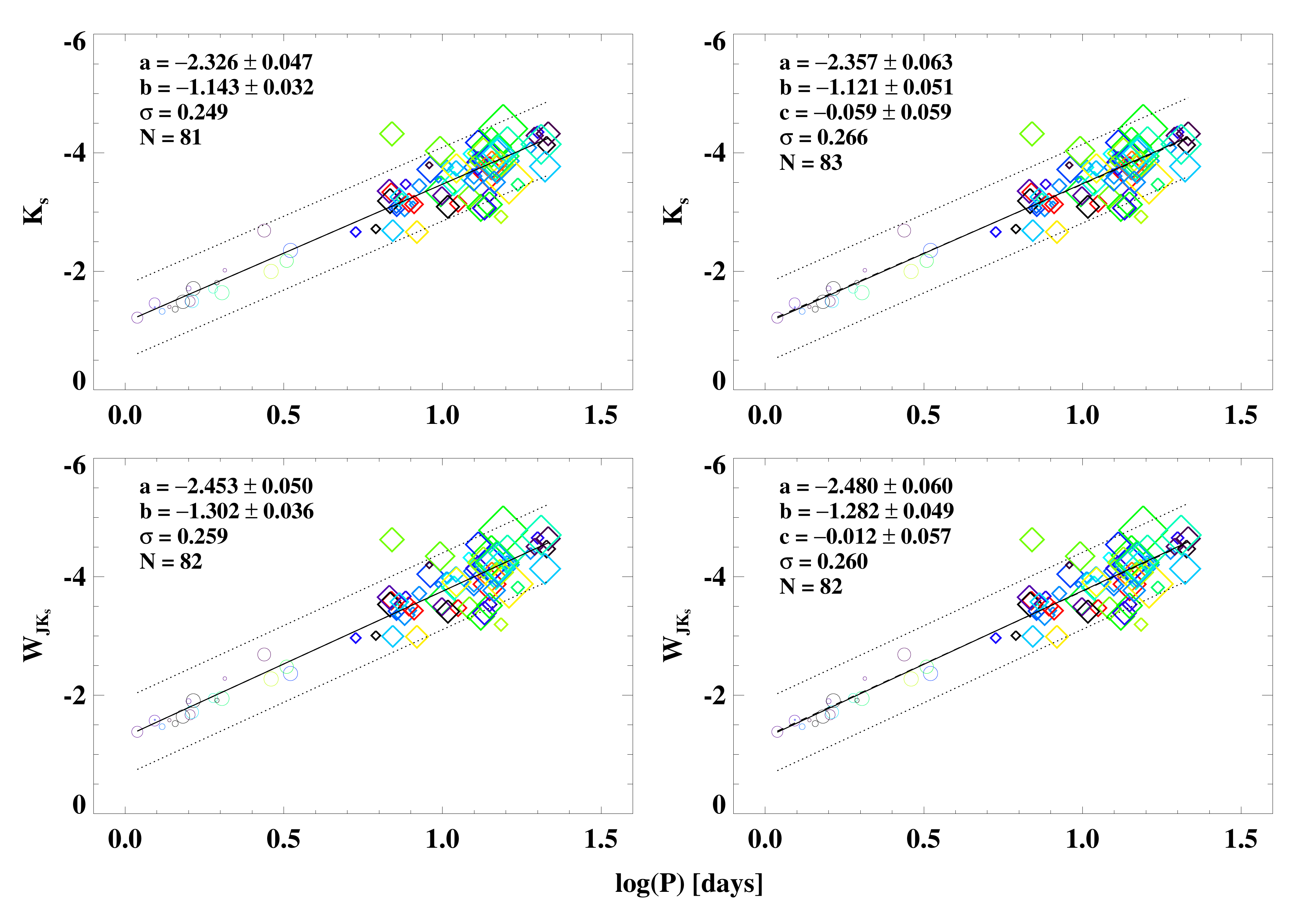}
    \caption{{\it Top:} $K_s$-band PL relation (left) and PLZ relation (right) for the combined sample of BL Her and W Vir stars. {\it Bottom:} Same as the top panels but for $W_{JK_s}$ Wesenheit magnitudes. The slope ($a$), zero-point ($b$), metallicity coefficient ($c$), dispersion ($\sigma$) and the number of Type II Cepheids in the final fit ($N$) are mentioned in each panel. Note that the symbol size represents the error in parallaxes while colors represent variation in [Fe/H] values. The solid lines show the best-fitting linear regressions and the dashed lines represent $\pm3\sigma$ scatter.}
    \label{fig:t2c_plr}
  \end{figure*}

\subsection{NIR Period-Luminosity-Metallicity relations}

Other than spectroscopic metallicities and accurate {\it Gaia} parallaxes, NIR light curves of Type II Cepheids are also needed to derive precise PLZ relations. However, NIR light curves are not available for our sample of 166 Type II Cepheids and are being obtained using the REM telescope, as are those of Classical Cepheids. Nevertheless, we can do a preliminary analysis using the random-phase 2MASS observations. Given the large uncertainties in parallaxes for RV Tau stars, we restricted our sample to BL Her and W Vir variables. It is known that the PL relations of RV Tau do not follow their short-period counterparts \citep{ripepi2014a, bhardwaj2017a}. Therefore, we excluded RV Tau stars and utilized a combined sample of 85 BL Her and W Vir stars for PLZ analysis.

Fig.~\ref{fig:t2c_plr} displays the PL and PW relations for Type II Cepheids in the left panels. The extinction corrections were applied assuming the reddening law of \citet{card1989}. The precise parallaxes of BL Her stars result in tighter PL/PW relations, but the scatter increases for W Vir stars. Note that the dispersion in these relations is nearly two times larger than the scatter in the PL relations for Type II Cepheids in globular clusters \citep{matsunaga2006, bhardwaj2022, ngeow2022a}. This increase in scatter can be attributed to random-phase 2MASS magnitudes and larger uncertainties in the parallaxes of W Vir stars. For these field stars with periods smaller than 10 days, the scatter in the PL relations is comparable to cluster variables. The right panels display PLZ relations after including the metallicity term. It is evident that the metallicity coefficient is statistically consistent with zero and this dependence is in agreement with the pulsation model predictions \citep{das2021}. Our ongoing optical ($griz$) and NIR ($JHK_s$) observations of Type II Cepheids using REM will allow us to constrain better the empirical metallicity coefficients at multiple wavelengths.

\section{RR Lyrae variables}

Similar to Classical Cepheids, one of the main issues in RR Lyrae-based distance determinations is the impact of metallicity dependence on their PL relations. While for Cepheids, these effects had previously been predicted to be smaller, the luminosities of RR Lyrae are known to depend significantly on their metallicities. \citet{catelan2004} predicted a metallicity coefficient of $\sim 0.2$ mag/dex in $JHK_s$ PL relations using models of horizontal branch evolution, and \citet{marconi2015} found similar dependence using RR Lyrae pulsation models. However, several empirical studies in the recent past predicted a small or negligible dependence of metallicity in NIR PL relations (see Table~\ref{tab:rrl}). To resolve these inconsistencies in empirical calibrations, we observed RR Lyrae variables in several globular clusters of different mean-metallicities at these longer wavelengths.

\subsection{NIR monitoring of RR Lyrae variables in globular clusters}

RR Lyrae variables in globular clusters provide a unique opportunity to study the variation in their pulsation properties as a function of metallicity, assuming that the clusters under consideration are mono-metallic or do not exhibit a significant metallicity spread. For this purpose, we started an observational program to monitor several globular clusters in $JHK_s$ bands using the WIRCam instrument at the Canada-France-Hawaii Telescope. To increase the metallicity baseline, we are also observing several metal-rich bulge clusters using the Flamingos 2 instrument at the Gemini-South Telescope. The observational program consisted of typically 10-15 epochs per globular clusters with each epoch separated by at least 2 hours. This ensured a full phase coverage for RR Lyrae light-curves in NIR bands. The goal was to quantify the metallicity term by comparing cluster RR Lyrae stars with metallicity differences of up to $\Delta$[Fe/H]$\sim 2$~dex, when the effect of metallicity difference on their luminosities ($\Delta M\sim 0.4$~mag) will be quite evident assuming the predictions of theoretical models.

  \begin{figure*}
    \includegraphics[scale=.54]{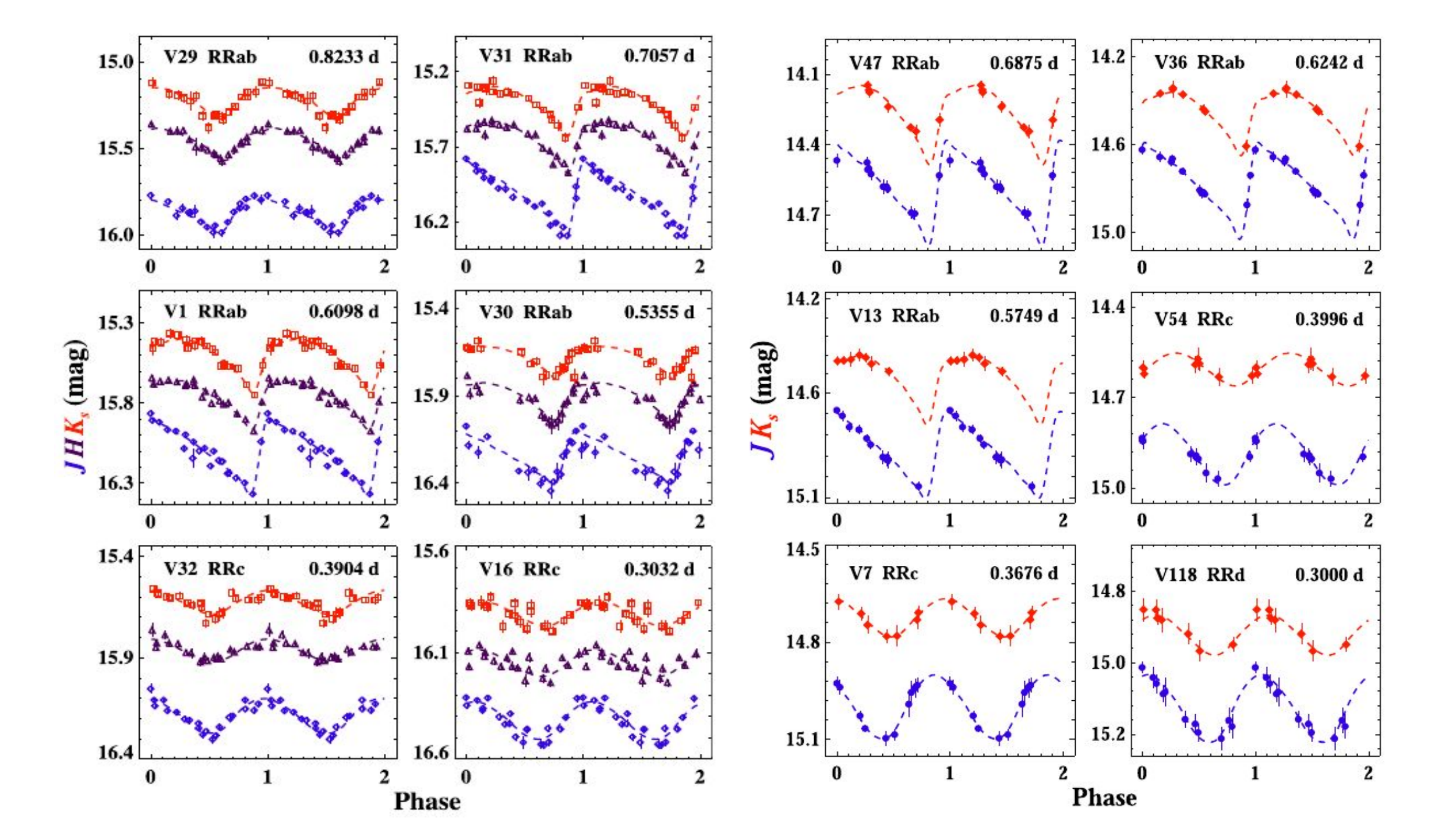}
    \caption{Example light curves for RR Lyrae stars in two different clusters - M53 (left) and M15 (right) with different light curve sampling \citep{bhardwaj2021, bhardwaj2021a}. The best-fitted templates are also shown as dashed lines. The magnitudes in different filters are slightly offset for visualization purposes.}
    \label{fig:rrl_lcs}
  \end{figure*}

Fig.~\ref{fig:rrl_lcs} displays example light curves of RR Lyrae variables with different periods in two clusters having different number of epochs per light curve. The left panels represent variables in M53 \citep{bhardwaj2021}, which are the best sampled with more than 20 epochs per light curve. The right panels display M15 variables with sparsely sampled light curves derived from archival imaging data \citep{bhardwaj2021a}. More epochs are now available for this cluster, which was observed with both CFHT and Gemini Telescopes in our observational program. The NIR light curves were fitted using the templates derived by \citet{braga2019}. While these templates were based on NIR light curves of RR Lyrae in $\omega$ Cen and were divided in three separate period bins for fundamental mode variables (RRab), they do not necessarily fit our light curves in other clusters in the given period-bin. This is expected since the period distributions of RR Lyrae stars vary in different clusters depending on their metallicities. We, therefore, fitted all three sets of templates for RRab stars to observed NIR light curves. For overtone mode (RRc) and mixed-mode (RRd) stars, the single set of template from \citet{braga2019} works well for all periods. Fig.~\ref{fig:rrl_lcs} shows the best-fitted light curve templates for RR Lyrae stars with different periods. For each RR Lyrae in a given cluster, we also obtained the mean-metallicity of the cluster from \citet{carretta2009} and the reddening value from the catalog of \citet{harris2010}.

\subsection{Color--magnitude diagrams and period-luminosity relations}

  \begin{figure*}
    \includegraphics[scale=.53]{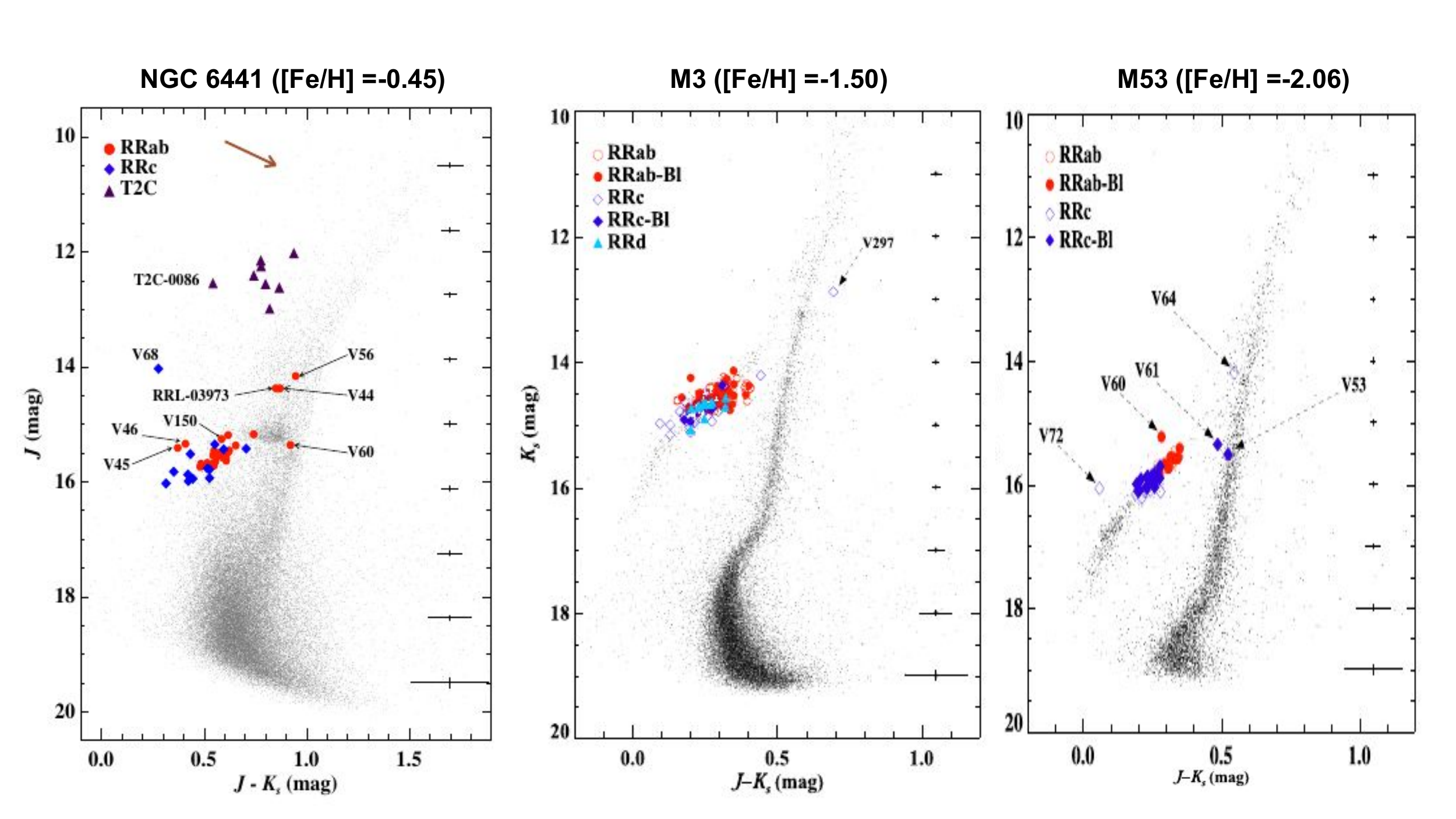}
    \caption{NIR color--magnitude diagrams for RR Lyrae variables in different globular clusters.}
    \label{fig:rrl_cmd}
  \end{figure*}
  
Template-fitted light curves of RR Lyrae variables in different globular clusters were used to determine their accurate and precise mean-magnitudes and colors. The extinction corrections were adopted using the \citet{card1989} reddening law together with $E(B-V)$ values from \citet{harris2010}.
Fig.~\ref{fig:rrl_cmd} displays color--magnitude diagrams for three different globular clusters with varying mean-metallicities. It is evident that these clusters are well-populated with RR Lyrae stars. NGC 6441 is a peculiar metal-rich cluster with red horizontal branch and hosts RR Lyrae stars. RR Lyrae variables in this cluster have unusually long periods and are considered to be helium enhanced \citep{pritzl2003, catelan2009}. However, our NIR observations of RR Lyrae variables in NGC 6441 for the first time showed that their NIR magnitudes and colors are more in agreement with stellar pulsation models adopting canonical helium content \citep{bhardwaj2022a}. RR Lyrae models with helium enhancement predict bluer colors and brighter magnitudes than observations. In the case of metal-intermediate M3 and metal-poor M53 \citep[see][for details]{bhardwaj2020, bhardwaj2021}, the magnitudes and colors of RR Lyrae stars fall within the predicted boundaries of the instability strip \citep{marconi2015} at NIR wavelengths. 

  \begin{figure*}
  \centering
    \includegraphics[scale=.32]{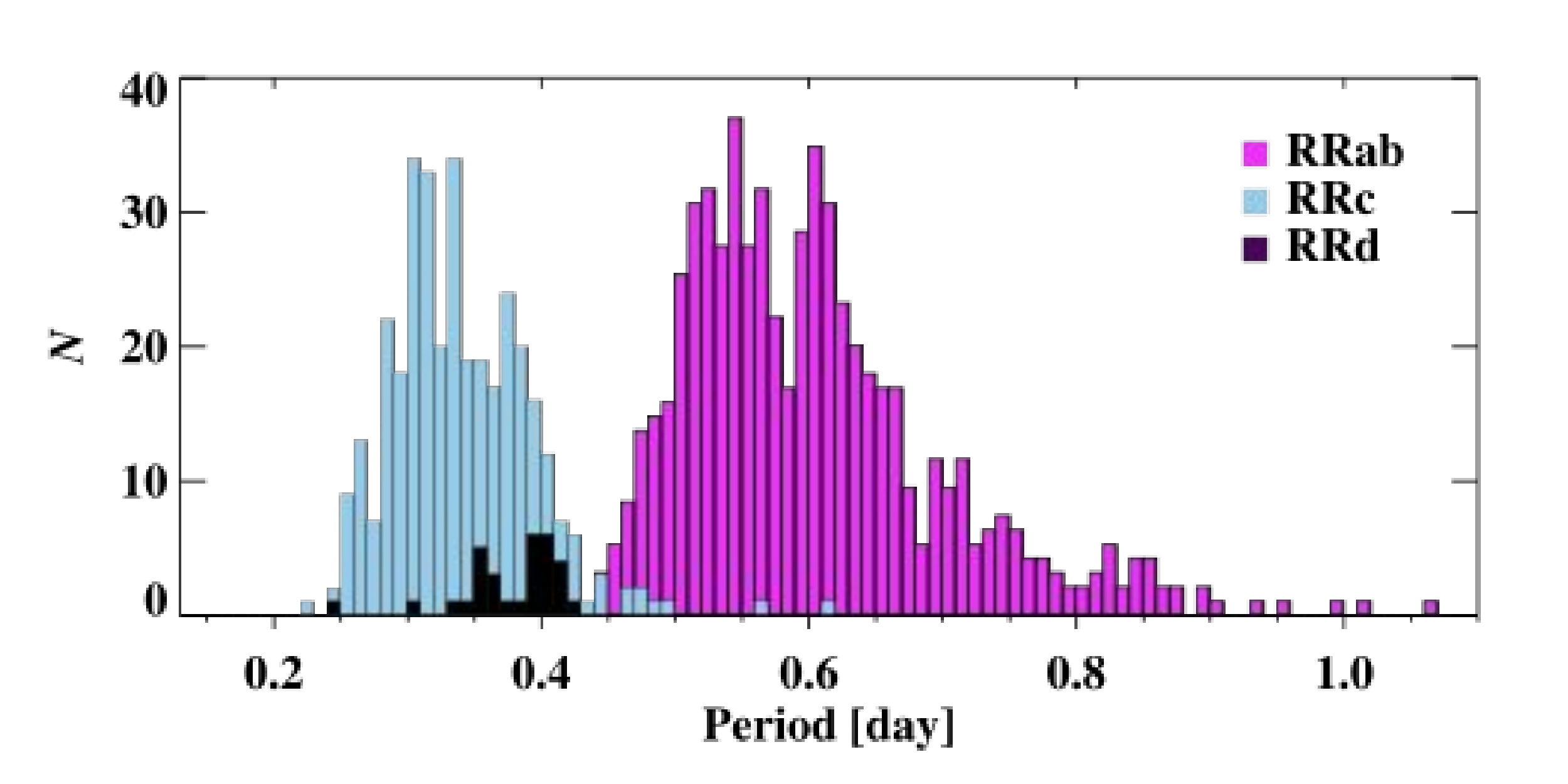}
    \caption{Histogram of period distributions of 964 globular cluster RR Lyrae variables.}
    \label{fig:rrl_per}
  \end{figure*}

We investigated PL relations for RR Lyrae stars in globular clusters with different metallicities. 
For this purpose, we utilized our homogeneous photometric data in several clusters: NGC 6441, NGC 6121, NGC 6402, NGC 5272, NGC 6934, NGC 7089, NGC 5024, and NGC 7078. We supplemented our clusters with already published data in the literature for NGC 5904, NGC 5139, and NGC 6341. Interested readers are referred to \citet{bhardwaj2023} for the details regarding these datasets. Fig.~\ref{fig:rrl_per} displays the histograms of period distributions of all 964 cluster RR Lyrae variables in our sample. 
We derived PL relations for different subtypes of RR Lyrae stars and the combined sample of RRab+RRc+RRd stars by fundamentalizing the periods of RRc/RRd stars using: $\log(P_{f})=\log(P_{fo})+0.127$ \citep{braga2022}. We assume that the RRd stars exhibit the first-overtone mode as the dominant mode of pulsation. 
  
  \begin{figure*}[!h]
    \includegraphics[scale=.37]{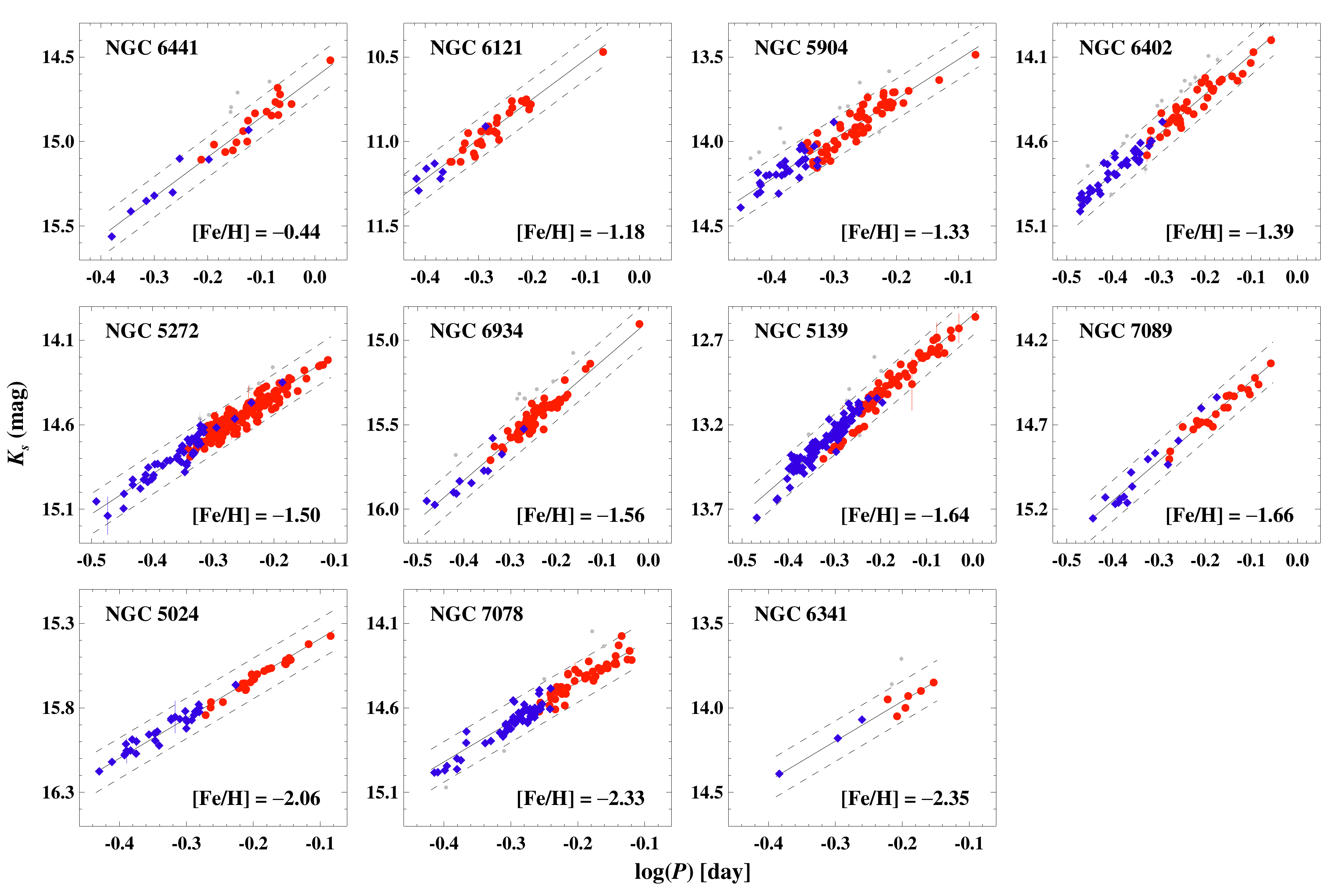}
    \caption{$K_s$-band PL relations for RR Lyrae variables in globular clusters with different mean-metallicities. The red circles and blue squares represent RRab and RRc stars respectively. The best-fitted linear regression is shown in solid lines while the dashed lines display $3\sigma$ scatter in the underlying relation. The small grey dots represent the outliers.}
    \label{fig:rrl_plr}
  \end{figure*}

Fig.~\ref{fig:rrl_plr} displays $K_s$-band PL relations for RR Lyrae variables in different globular clusters, and we have similar relations in $JH$ bands as well \citep{bhardwaj2023}. In the case of metal-rich NGC 6441, NIR PL relations are tight (with dispersion of $\sim0.05$~mag) and fully consistent with the predictions of the pulsation models with canonical helium content from \citet{marconi2018}. The predicted PL relations with higher helium content result in systematically brighter zero-points even though the slopes remain statistically consistent with observations. Therefore, we concluded that RR Lyrae variables in NGC 6441 are either not significantly helium enhanced, as previously thought, or the impact of such enhancement in NIR bands is much smaller than the predictions of the pulsation codes \citep{bhardwaj2022a}. In the case of normal metal-intermediate and metal-poor clusters, the PL relations in NIR agree well with theoretically predicted relations by \citet{marconi2015}. For high latitude clusters with less extinction and field contamination, the PL relations for RR Lyrae stars are very tight. For example, the dispersion of $\sim0.02$~mag for RRab in M53 is at the limit of photometric uncertainties. The bulge clusters with severe crowding and differential extinction result in relatively larger scatter (with dispersion of $\sim0.07$~mag) in the PL relations. Nevertheless, the slopes of PL relations in different clusters are statistically consistent for a given sample of RR Lyrae subtype. The PL relations for the combined sample of all RR Lyrae stars are best constrained thanks to the larger statistics and longer period range under consideration.

\subsection{NIR period-luminosity-metallicity relations}

  \begin{figure*}
    \includegraphics[scale=.55]{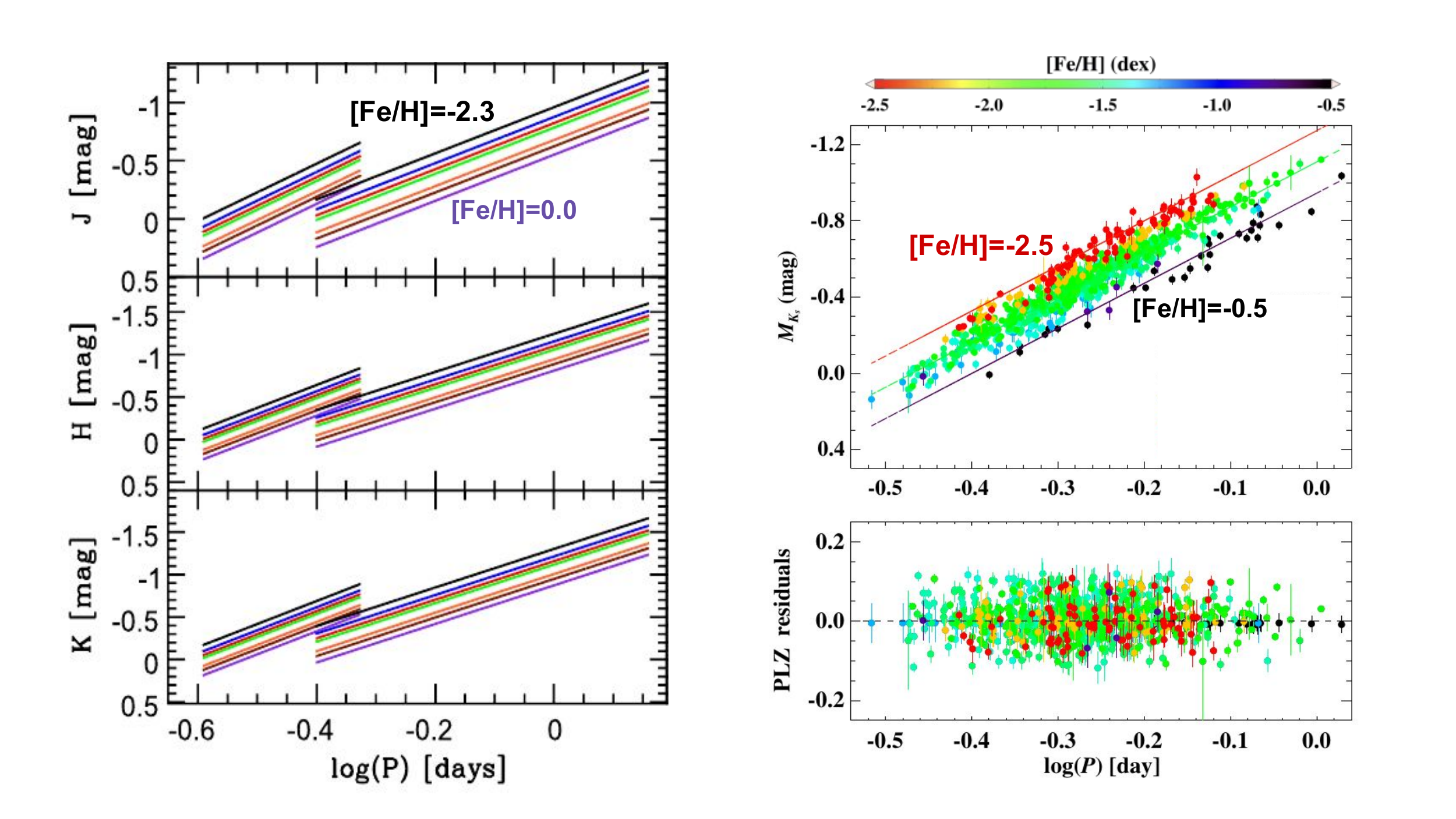}
    \caption{{\it Left:} NIR PLZ relation for RR Lyrae variables predicted by \citet{marconi2015} based on stellar pulsation models. The figure is adopted and slightly modified from the original Figure 14 of \citet{marconi2015}. {\it Right:} Empirical PL relations for RR Lyrae stars in different globular clusters as a function of metallicity (top). The bottom panel displays the residuals of PLZ relation after taking into account the metallicity term. It is evident that the trend in the residuals as a function of metallicity, as seen in the top panels, vanishes in the lower panel. }
    \label{fig:rrl_plz}
  \end{figure*}

We used individual PL relations in different globular clusters to investigate and quantify the dependence of metallicity at NIR wavelengths. Previously, \citet{sollima2006} used random-phase $K_s$-band photometry for more than 500 RR Lyrae variables in 16 globular clusters covering a metallicity range of $\sim1.2$ dex to derive PLZ relation. For our analysis, homogeneous NIR photometry for 964 RR Lyrae variables in 11 globular clusters covering $\Delta$[Fe/H]$\sim2$~dex was combined with NIR photometry of 346 RR Lyrae stars in the Milky Way field. The sample of field 
RR Lyrae variables was adopted from \citet{muraveva2018} with metallicities and reddening estimates. NIR photometry for these variables was adopted from 2MASS catalog as discussed in \citet{bhardwaj2023}. These field stars with accurate parallaxes from {\it Gaia} allowed us to constrain an absolute zero-point of the PLZ relation in $JHK_s$ bands. We derived a PLZ relation for the combined sample of 1310 field and globular cluster RR Lyrae variables by simultaneously solving for a slope, zero-point, metallicity coefficient, and individual distances to globular clusters. This approach does not depend on prior assumption of distance to a given globular cluster. 

The left panels of Fig.~\ref{fig:rrl_plz} display theoretical PLZ relation in $JHK_s$ band for RR Lyrae variables as predicted by the pulsation models. The right panel displays our empirical relation in $K_s$-band and similar relations are also derived in $JH$ filters as well. The combined sample of all RR Lyrae stars was used in deriving these PLZ relations.
The variation in the PL relations as a function of metallicity is clearly evident in both the theoretical and empirical relations. We find an excellent agreement between the metallicity coefficients from the theoretical models and the observations. To investigate the possible effects of different metallicity scales and reddening laws, we also adopted recently published spectroscopic metallicities of field stars by \citet{crestani2021}. For the globular clusters, we experimented with different metallicity scales of \citet{carretta2009}, \citet{harris2010}, and \citet{dias2016}. No statistically significant variation was seen in the PLZ relations for different metallicity scales. Similarly, extinction corrections based on different reddening maps \citep[][]{schlafly2011, green2019} resulted in small or negligible variations in PLZ coefficients within their associated uncertainties.

Our recommended PLZ relations in NIR filters are:
\begin{eqnarray}
   ~~~~~~~~M_J =-0.44~(\pm0.03) -1.83~(\pm0.02)\log(P_\textrm{f}) + 0.20~ (\pm0.02)~\textrm{[Fe/H]}~~(\sigma=0.05~\textrm{mag}), \nonumber \\ 
  ~~~~~~~~M_H =-0.74~(\pm0.02) -2.29~(\pm0.02)\log(P_\textrm{f}) + 0.19~ (\pm0.01)~\textrm{[Fe/H]}~~(\sigma=0.05~\textrm{mag}), \nonumber \\ 
  ~~~~~~~~M_{K_s} =-0.80~(\pm0.02) -2.37~(\pm0.02)\log(P_\textrm{f}) + 0.18~ (\pm0.01)~\textrm{[Fe/H]}~~(\sigma=0.05~\textrm{mag}), \nonumber  \\ \nonumber
\end{eqnarray}
\noindent where $M_\lambda$ represents the absolute magnitude in the $JHK_s$ filters, and $P_f$ represents the periods of RRab stars or the fundamentalized periods of RRc/RRd stars. These empirical relations are fully consistent with the theoretical relations provided by \citet{marconi2015}. Our empirical calibrations will be used to determine robust RR Lyrae based distances to nearby dwarf spheroidal galaxies like Draco, Sextans, and Ursa Minor, for which NIR photometry using WIRCam at CFHT has already been obtained. These precise and accurate RR Lyrae distances will then allow us to provide an independent calibration of other population II distance indicators such as the tip of the red giant branch for the population II distance scale.

Since we solved for the individual distances to our sample of globular clusters, we can compare those with independent distances in the literature. We adopted globular cluster distances from the catalog of \citet{baumgardt2021} which provides average distances based on several independent determinations. A direct comparison of our distance estimates with those provided by \citet{baumgardt2021} suggests a small offset of $-0.03\pm0.03$~mag such that our distances are systematically smaller albeit within $1\sigma$ errors. This difference can be explained due to overcorrection of parallax zero-point offset, because a smaller parallax correction than \citet{lindegren2021} will result in brighter absolute zero-points \citep{bhardwaj2021} and in turn a larger distance to our sample of globular clusters. Nevertheless, our independently derived RR Lyrae based distances also agree well with those available in the literature within their quoted uncertainties.

\section{Summary}

Classical pulsators, in particular, Classical Cepheids, Type II Cepheids, and RR Lyrae variables are stellar standard candles that are used to obtain accurate and precise distances to their host galaxies. The advent and the increase of NIR detectors in the past two decades have improved the accuracy and precision of these variables as distance indicators since their PL relations are tighter at these longer wavelengths. Despite the extensive use and improvements in the PL relations, their universality in different stellar systems is not well-established. In particular, the metallicity effects on NIR PL relations for Cepheid and RR Lyrae variables are not properly quantified empirically and there are inconsistencies between observations and theoretical model predictions. 

We discussed our ongoing efforts to properly quantify the metallicity effects on the PL relations for these classical pulsators by utilizing the increasingly accurate {\it Gaia} parallaxes. For the precise calibration of these relations, we also need high-resolution spectroscopic metallicities and multiband time-series photometry for a statistically significant sample of variables. Our observational programs are obtaining homogeneously both the spectra and the light curves for these classical pulsators to fully utilize current and future {\it Gaia} parallaxes. We presented preliminary results on the high-resolution spectroscopic metallicities of Milky Way Cepheid standards that have been used in the latest local determination of the Hubble constant. Our new spectroscopic metallicities together with time-series photometry and {\it Gaia} parallaxes will provide a new calibration for the distance scale. At the same time, the C-MetaLL survey is obtaining spectra for most metal-poor Classical Cepheids in our Galaxy to increase the metallicity baseline for a better quantification of metallicity effects. For Type II Cepheids, we show that NIR light curves are equally important when deriving PLZ relations even with a homogeneous sample of spectroscopic metallicities and accurate {\it Gaia} parallaxes. For RR Lyrae stars, we presented the most precise quantification of metallicity dependence to date using variables in globular clusters of different mean-metallicities. Our latest results and ongoing efforts to improve the accuracy and precision of PL calibrations will be useful for distance scale studies in the near-future. These empirical calibrations will be particularly important in the era of upcoming large ground-based telescopes and space telescopes, which will operate mostly at longer wavelengths.\\

{\it This project has received funding from the European Union's Horizon 2020 research and innovation program under the Marie Sklodowska-Curie grant agreement No. 886298.}

\bibliographystyle{mnras}
\bibliography{mybib_final.bib}

~\\
~\\
\noindent {\bf Discussion}\\
\vspace{-6pt}

\noindent {\bf Question (Braga):} Did you check what happens to the matrix solution when removing $\omega$ Cen RR Lyrae, since for them, you adopted the average [Fe/H] value of $\omega$ Cen, but this globular cluster is known for having a large iron abundance spread, even among its RR Lyrae stars.\\
\noindent {\bf Answer:} Thanks. We did check this in \citet{bhardwaj2021} by adopting different metallicities of $\omega$ Cen RR Lyrae. For this work, I have not checked this, but I doubt this will make any difference since the overall metallicity range is significantly larger.\\

\noindent {\bf Question (Riess):} Very nice talk. What did you conclude from, or did you, compare RR Lyrae distances for globular clusters to {\it Gaia} distances? Did you sort of average stars in {\it Gaia}?\\
\noindent {\bf Answer:} We didn’t compare directly with {\it Gaia} because some of these clusters are quite distant. We have a few of these clusters at 15-19 kpc where {\it Gaia} parallaxes are unlikely to be that good. But we did compare with some independent measurements based on, for example, the compilation by \citet{baumgardt2021} where we did find a small offset, which is somewhat statistically consistent with zero.\\

\noindent {\bf Question (Anderson):} Given that Classical Cepheids and RR Lyrae stars occupy the same instability strip and share many similarities, what could be the origin of the different sign of the metallicity term for RR Lyrae and Cepheid PL-relations? For context, all individual PL relations and all PW relations of RR Lyrae stars shown in your talk have positive signs. Conversely, the observed metallicity effect sign for Cepheids is negative in both PL and PW relations, and models predict a negative sign for PW relations (and Geneva models also for individual bands). I wonder what physics this could point to.\\
\noindent {\bf Answer:} I have no idea. I think first we have to be sure that for Cepheids, it goes the other way. From the models, we see something else; from the observations we see something else. At least for RR Lyrae stars, I am now confident that we have consistency between models and observations. Having thought about it, RR Lyrae and Cepheids have different evolutionary parameters, mass-luminosity relations, compositions, and it will be difficult to speculate what might be contributing to different metallicity dependence.\\

\noindent {\bf  Comment (Hocd\'{e}):} Just to answer the question, I have a curious idea that needs to be checked – but this metallicity effect difference with Cepheids could come from circumstellar envelopes of Cepheids, if the emission of the envelopes depends on the metallicity. That could possibly be the physical effect that causes this. I can give details later.\\
\noindent {\bf Answer:} I think then one has to ask how much these effects are there at longer wavelengths or if there is a variation in the metallicity coefficients going from optical to infrared. I think this effect gets minimized as you go for longer wavelengths.\\

\noindent {\bf  Comment (Hocd\'{e}):}  For example, there is an interesting phenomenon which has to do with the opacity of the negative hydrogen ion which in fact depends on the quantity of the metals in the gas. I can also give details later. We have to study the physics of the negative hydrogen opacity. \\
\noindent {\bf Answer:} Sure, thanks.\\

\noindent {\bf Question (Martínez-V\'{a}zquez):} Very interesting talk. Have you considered (or tried already) to obtain period-$\phi$31-[Fe/H] relations in the different infrared bands you have?\\
\noindent {\bf Answer:} Not yet, but certainly in the list things to do.  This is something I thought about when the RRab templates from \citet{braga2019} did not work well based on their period bins, since the shape of RRab light curves varies at different periods in different metallicity clusters.\\

\end{document}